# Broadband Reflective Elastic Mode Conversion Enabled by a Single Row of Inclined Long-Slits

Kaifei Feng[1], Weidong Wang[1], Yucheng Gao[2], Fengming Liu, Qiujiao Du, Wenshuai Zhang, and Pai Peng[1,*]

[1]School of Mathematics and Physics, China University of Geosciences, Wuhan 430074, China

[2]School of Mechanical and Manufacturing Engineering, University of New South Wales, Sydney, Australia

*Corresponding author: paipeng@cug.edu.cn

## Abstract

Broadband longitudinal-to-transverse mode conversion under normal incidence remains difficult to achieve, especially with structurally simple designs. Numerical simulations show that a single periodic row of inclined long-slits near a free surface enables high-efficiency broadband conversion, where the conversion rate exceeds 0.8 across a normalized-frequency range of 0.27–0.74, corresponding to a 93.1% relative bandwidth. The broadband response originates from two intrinsic deformation modes of the mass blocks between adjacent inclined slits: a rotational mode and a quadrupole mode. The spectral overlap of these two modes sustains a continuous high-efficiency band. The effect is robust against geometric variations, demonstrating that geometric asymmetry alone offers a minimal yet effective route to broadband elastic-wave mode conversion.

In recent years, the development of elastic metamaterials has opened new possibilities for elastic-wave manipulation[1-13] . The wave control functions mainly focus on the regulation of amplitude[1], phase[2], and polarization mode[3-13] of elastic waves. In particular, mode conversion between longitudinal (L) and transverse (T) waves plays a central role in polarization engineering, surface-wave control, and energy localization[14-16]. Elastic mode conversion fundamentally requires breaking the system symmetry so that L and T motions, which are orthogonal in isotropic media, can become coupled. A classical way to introduce such asymmetry is to use oblique incidence, including well-known effects such as Brewster-angle conversion. Unfortunately, many application scenarios require a plane wave arriving from infinity under normal incidence. Kweun *et al* [4]first demonstrated normal-incidence mode conversion by employing effective anisotropic media to achieve total mode conversion. After that work, many other mode-conversion schemes were proposed. Wang *et al.* [5] realized nearly total normal-incidence L-to-T conversion using an ultrathin elastic metamaterial plate with oblique anisotropic dipolar resonances . Clark *et al.*[6] demonstrated mode conversion using a meta-structure that rotates the displacement field through curved wave-pipe guides. More recently, Chai *et al.*[7] reported full-power mode conversion using a bilayer metastructure formed by two arrays of periodic slits that produce asymmetric transmission.

So far, achieving high conversion efficiency over a broad frequency range remains a challenge. If "high efficiency" is defined as a conversion rate exceeding 0.8, most reported designs [4-12] achieve high conversion only over narrow bands, typically below 10% relative bandwidth, as exemplified by the transmodal Fabry–Pérot resonances [4]. To date, the only design (to the best of our knowledge) that achieves a relatively broad conversion band is the machine-learning-assisted structure reported in Ref.[13], where a relative bandwidth of about 79% was obtained. That approach represents a powerful data-driven route to mode-conversion engineering. Nevertheless, its geometry is highly complex, and its large void fraction ($\approx$50%) limits its suitability for applications requiring structural integrity. In this work, we explore an alternative route and achieve a relative bandwidth of 93.1% using only a single periodic row of inclined long-slits placed near a free surface. The design does not rely on machine-learning optimization, gradient structures, or multilayered configurations. Its broadband response can be interpreted in terms of two local deformation modes excited by the slit inclination: a rotational mode and a quadrupole mode. The spectra of these modes lie close to each other, and their combined excitation produces a continuous high-efficiency conversion band.

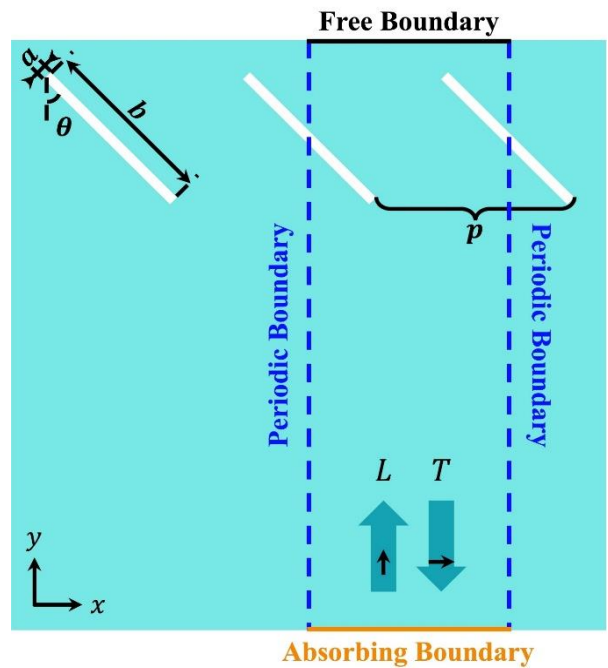

Fig 1. Schematic of the two-dimensional half-space elastic system. The dashed box indicates the calculation region. The top and bottom are free and absorbing boundaries, respectively. The left and right are periodic boundary.

We consider a two-dimensional system consisting of a semi-infinite epoxy domain, as shown in Fig. 1. The upper boundary is the interface, serving as a free surface. A periodic array of inclined long-slits with a period $p$ is embedded near this surface. Each slit has a width $a = 0.02p$, a length $b = 0.82p$, and a rotated angle $\theta = 45°$ (respect to vertical direction). The material parameters are $\rho_e = 1180 \text{kg/m}^3$, $c_{l,e} = 2540 \text{m/s}$, and $c_{t,e} = 1160 \text{m/s}$ for epoxy, with $\rho_a = 1.29 \text{kg/m}^3$ and $c_a = 343 \text{m/s}$ for air. The mode conversion behavior is analyzed using a scattering model, where the computational region is denoted by the dashed box in Fig. 2. In the absence of dissipation, elastic plane waves are normally incident from the bottom and fully reflected by the top free surface. We focus on the conversion of incident L waves into reflected T waves. The L-to-T conversion rate (CR) is defined as the ratio of the vertical energy fluxes of the two waves. The CR ranges from 0 to 1 as dictated by energy conservation and is calculated numerically using the finite-element method implemented in COMSOL Multiphysics.

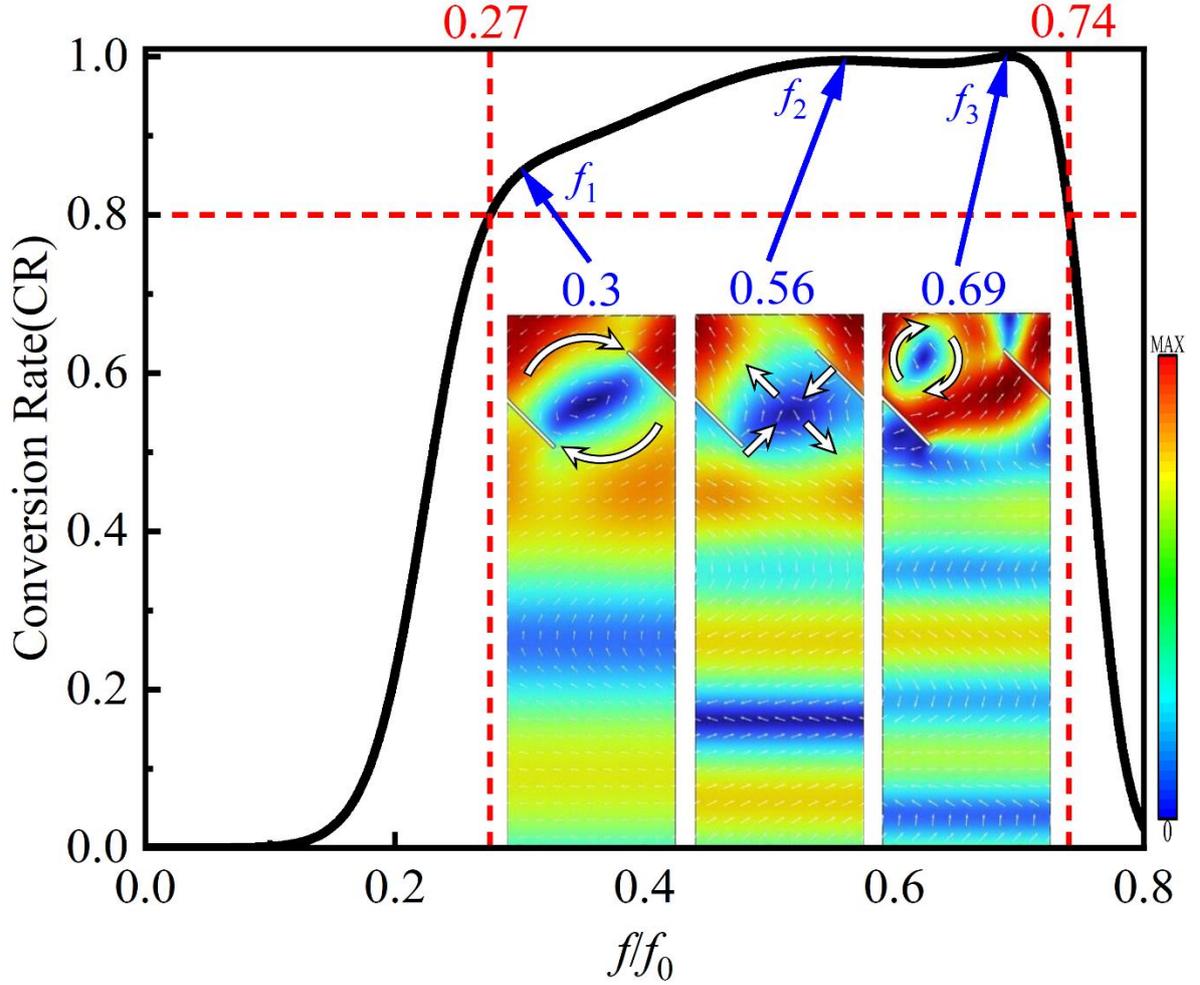

Fig 2. Conversion rate (CR) varies with frequency $f$. normalized by $f_0 = 1160 m/s$. Three representative excitation fields (insets) at $f/f_0$ = 0.3, 0.56, and 0.69 illustrate distinct conversion modes that collectively yield a broadband high-efficiency response. Color denotes the displacement magnitude, and arrows indicate the displacement direction.

Figure 2 presents the CR as a function of frequency $f$ (normalized by $f_0 = c_t / p$). The proposed inclined long-slits structure exhibits outstanding high performance: the CR remains above 0.8 throughout the range $0.27 < f/f_0 < 0.74$ (marked by red dashed lines), The relative bandwidth, defined as $2(f_H - f_L)/(f_H + f_L)$ [17], is about 93.1%, where $f_H$ and $f_L$ correspond to the upper and lower bound frequencies respectively. Within this wide band, three distinct peaks appear at $f_1 = 0.3f_0$, $f_2 = 0.56f_0$, and $f_3 = 0.69f_0$ (marked by blue arrows), whose spectral overlap forms a continuous broadband plateau. The displacement fields shown in the insets of Fig. 2 illustrate the motion patterns at the peaks. The patterns at $f_1$ and $f_3$ are dominated by the rotational motion of the mass block between adjacent inclined slits, whereas the mode at $f_2$ exhibits quadrupolar motion. The thick white arrows indicate the overall vibration direction of each mode.

These three motion patterns are further validated by the eigenfield analysis. Figure 3(a) shows the band structure of the long-slit unit cell, where three eigenmodes (mode A, mode B, and mode C) are identified at the $\Gamma$ point. Their eigenfrequencies ($f_A = 0.29f_0$, $f_B = 0.62f_0$ and $f_C = 0.71f_0$) closely match the three conversion peaks ($f_1 = 0.3f_0$, $f_2 = 0.56f_0$, and $f_3 = 0.69f_0$). The small differences may stem from using an isolated unit cell with free top and bottom boundaries in the band-structure calculation, while the conversion analysis attaches the until cells to an epoxy half-space, where substrate coupling modifies their frequencies and fields. The corresponding eigenfield distributions in Figs. 3(b–d) closely resemble the displacement fields at the corresponding conversion peaks. In particular, the eigenfield of mode C reveals a second-order rotational motion of the inter-slit mass block. The agreement between the eigenfields and the excited fields shows that the broadband response can be traced to the cooperative excitation of several local deformation modes. These modes act together to sustain a continuous and efficient mode-conversion band. The rotational and quadrupolar motions, which involve horizontal displacements, cannot be excited by normally incident L waves that contain only vertical displacements because of the system symmetry. The inclined slits break this symmetry and "tilt" the local vibration axes, causing these rotational and quadrupolar motions to become obliquely. As a result, the oblique vibrations couple the horizontal and vertical displacements, leading to elastic mode conversions.

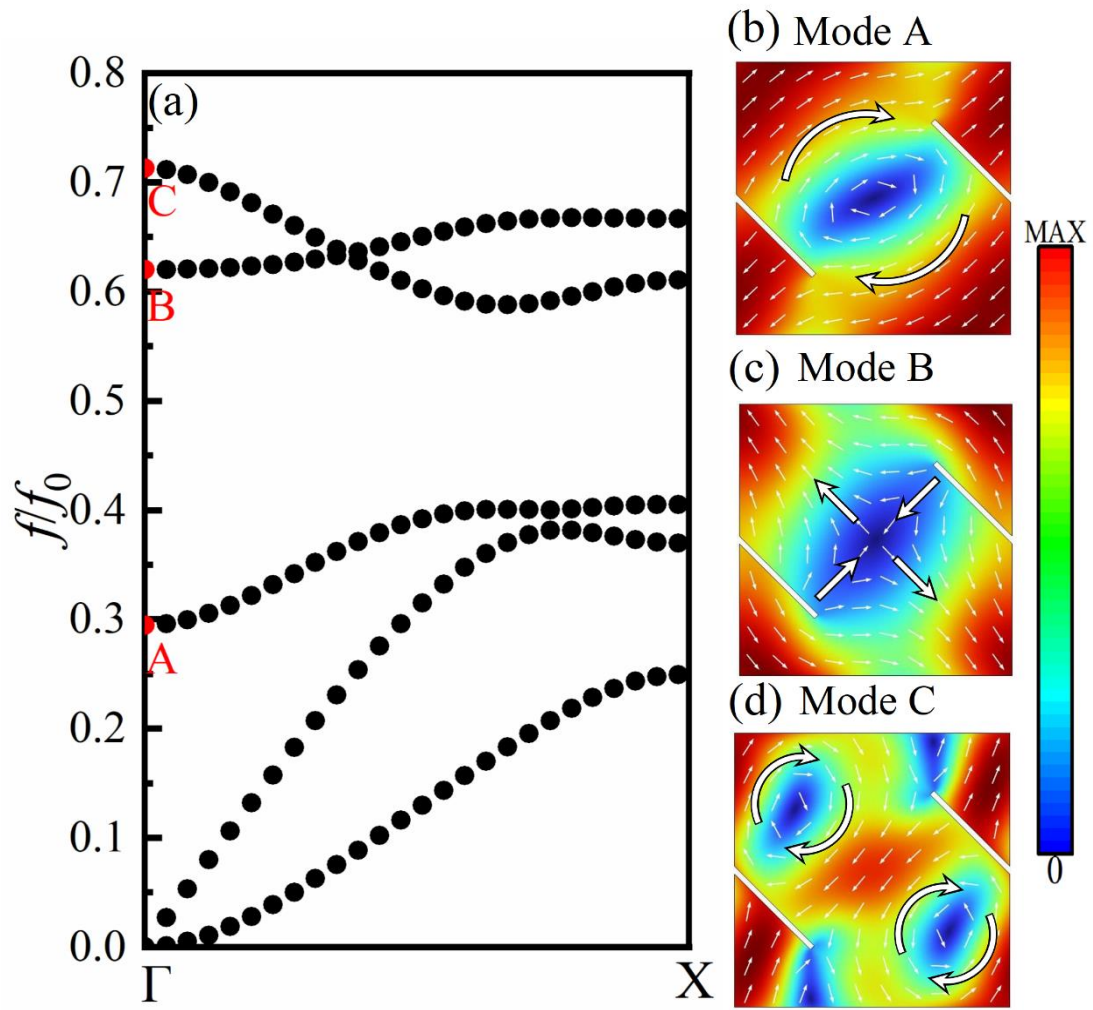

Fig 3. (a) The band structure of a square long-slit unit cell with side length $p$. The top and bottom boundaries are free, while the left and right boundaries are periodic. (b) (c) and (d) are the displacement eigenfields for the mode A, B, and C, respectively.

To our knowledge, these rotational and quadrupolar deformation modes discussed above have never been emphasized in studies of elastic wave mode conversions. Previous works have mainly employed symmetric structures, typically using slits arranged either vertically or horizontally [18, 19]. In our design, introducing a general rotated angle $\theta$ breaks this symmetry. The mode-conversion behavior is highly dependent on this key parameter, which provides a new degree of freedom. Figure 4(a) shows the CR spectra varying with rotated angle $\theta$. When the slits are vertical ($\theta = 0°$) or horizontal ($\theta = 90°$), modes A, B, and C still exist, but no conversion occurs because these rotational and quadrupolar motions cannot be excited by normally incident L waves. For small angles such as $\theta < 15°$ (or $\theta > 85°$), these three modes are weakly excited, leading to insufficient coupling and consequently low conversion efficiency. High conversion (CR > 0.8) appears when $\theta$ exceeds about $\theta \approx 40°$ and persists up to $\theta \approx 52°$. As $\theta$ increases within this range, the eigenfrequency of mode B shifts downward from about $0.7f_0$ to $0.5f_0$ and approaches that of mode A, causing the initially continuous broadband response to split into two partially separated peaks. Overall, a robust broadband conversion is maintained for $40° < \theta < 52°$, indicating that the broadband effect is stable against variations in the slit angle.

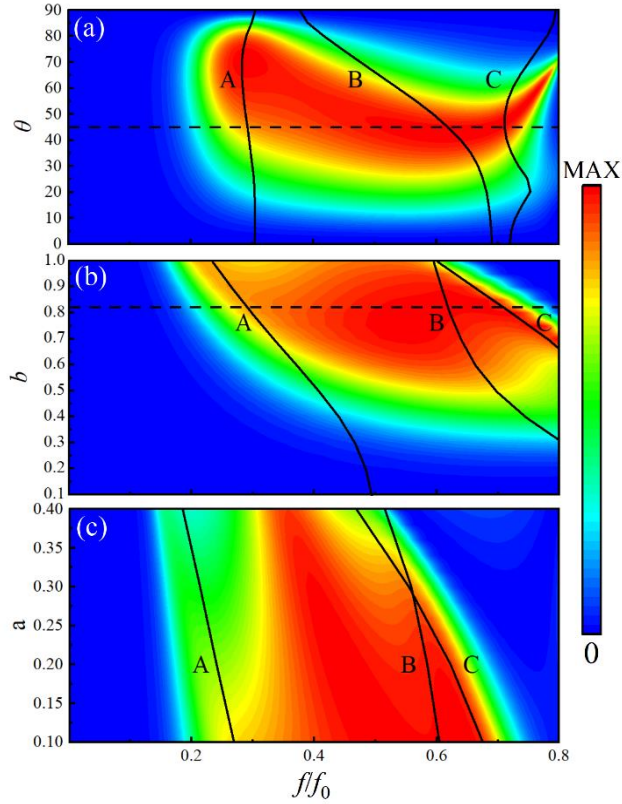

Fig 4. CR varying with the (a) rotated angle $\theta$, (b) length $b$ and (c) width $a$. Solid black lines indicate the evolution of the eigenfrequencies of modes A, B, and C (introduced in Fig. 3) as their parameters vary. Dashed black lines mark the parameter values corresponding to the results shown in Fig. 2.

We next turn to the geometric parameters. When the slit is very short ($b < 0.35$), the conversion remains weak over the entire frequency range. In this regime, the structure effectively behaves as a point-like scatterer that introduces only local perturbations and cannot sustain the multi-mode coupling required for broadband conversion. As the slit becomes sufficiently long ($b > 0.5$), the local vibration modes are tilted and a pronounced broadband region appears. During the elongation of the slit, the high-CR region gradually shifts toward lower frequencies. This trend can be interpreted as a geometrical scaling effect: extending the slit increases the characteristic length scale of the structure, which favors longer wavelengths and consequently red-shifts the operating band. Figure 4(c) shows that increasing the slit width gradually weakens the broadband conversion. As the slits widen, the three vibration modes A, B, and C are all suppressed because the mass blocks between adjacent inclined slits become increasingly compressed. When these mass blocks are almost squeezed out, the local rotational and quadrupolar modes associated with their motion disappear, and the broadband conversion vanishes accordingly. Overall, broad high-conversion bands persist within a wide range of $a$ and $b$, demonstrating the stability of the broadband effect.

In this work, we introduced a simple yet highly effective structure for broadband elastic-wave mode conversion. A single periodic row of inclined long-slits placed near a free surface is sufficient to produce efficient L-to-T conversion across a wide frequency range. The broadband behavior arises from the deformation dynamics of the local mass blocks created by the long-slits. Their intrinsic modes acquire mixed L and T displacement components once the slits are inclined, because the tilted geometry breaks the high symmetry of the system and reorients the vibration axes of these local motions. When the geometric parameters are chosen appropriately, several of these tilted local modes are excited with comparable strength. Their individual conversion peaks occur at nearby frequencies and merge into a continuous and high-efficiency broadband response.